# Exchange bias in phase-segregated $Nd_{2/3}Ca_{1/3}MnO_3$ as a function of temperature and cooling magnetic fields

Elena Fertman[1], Sergiy Dolya[1], Vladimir Desnenko[1], L.A. Pozhar[2], Marcela Kajňaková[3], and Alexander Feher[3]

[1]B. Verkin Institute for Low Temperature Physics and Engineering, NASU, 47 Lenin Ave., Kharkov 61103, Ukraine

[2]PermaNature, LLC., 1900 Corporate Drive, P.O. Box 380534, Birmingham, AL 35238, U.S.A.

[3]P. J. Šafárik University in Košice, Faculty of Science, Park Angelinum 9, 04154 Košice, Slovakia

## Abstract

Exchange bias (EB) phenomena have been observed in $Nd_{2/3}Ca_{1/3}MnO_3$ colossal magnetoresistance perovskite below the Curie temperature $T_C \sim 70$ K and attributed to an antiferromagnetic (AFM) – ferromagnetic (FM) spontaneous phase segregated state of this compound. Field cooled magnetic hysteresis loops exhibit shifts toward negative direction of the magnetic field axis. The values of exchange field $H_{EB}$ and coercivity $H_C$ are found to be strongly dependent of temperature and strength of the cooling magnetic field $H_{cool}$. These effects are attributed to evolution of the FM phase content and a size of FM clusters. A contribution to the total magnetization of the system due to the FM phase has been evaluated. The exchange bias effect decreases with increasing temperature up to $T_C$ and vanishes above this temperature with disappearance of FM phase. Relaxation of a non-equilibrium magnetic state of the compound manifests itself through a training effect also observed while studying EB in $Nd_{2/3}Ca_{1/3}MnO_3$.

## 1. Introduction

Exchange interaction at an interface between antiferromagnetic (AFM) and ferromagnetic (FM) materials causes additional unidirectional anisotropy of the magnetization, thus enabling a physical mechanism behind an important phenomenon called exchange bias (EB) [1, 2, 3]. The phenomenon manifests itself as a shift of a magnetic hysteresis loop along the magnetic field axis for the system cooled down in an external magnetic field below the magnetic ordering temperatures.

The EB phenomenon has been recently widely studied because of its importance in spintronic applications. The majority of experimental works in this field is focused on investigations of specially prepared systems, such as core-shell nanoparticles with the cores coupled to the shells [1, 4], or thin films composed of coupled FM and AFM layers [2, 5, 6, 7]. In parallel, theoretical models of EB in single crystal antiferromagnets with ideal [8] or non-ideal [9, 10, 11] interfaces were developed to serve as a common theoretical approach to quantification of EB effects. Later on, more complicated theoretical models applicable to EB in polycrystalline AFM-FM samples were also introduced [12].

The EB effect was observed not only in specially prepared systems, but also in bulk systems with spontaneous phase segregation. At high temperatures such compounds exist as single-phase systems, while becoming phase segregated (that is, having coexisting different magnetic phases) at low temperatures. Thus, the EB effect has been reported in mixed Laves phase compounds $RAl_2$ (R = Nd, Ho, Gd, etc.) [13, 14] and binary alloys, such as Ni-Mn [15], Fe-Mn [16], and Mn rich Heusler alloys [17, 18, 19].

In recent years the EB effect has been reported in perovskite manganites [20, 21, 22, 23] and cobaltites [24, 25, 26, 27], and attributed to their magnetic phase segregated states. In particular, at low temperatures a spontaneously phase segregated $Pr_{2/3}Ca_{2/3}MnO_3$ is composed of FM nanodomains immersed in a charge-ordered AFM host [20]. Exchange coupling at interfaces between the FM domains and the surrounding AFM matrix causes unidirectional anisotropy of the magnetization resulting in the EB effect. It is important to note, that phase separation is known to enable the colossal magnetoresistance effect in manganites [28]. To understand magnetic properties of such systems, more experimental and theoretical research is necessary.

The colossal magnetoresistive perovskite $Nd_{2/3}Ca_{1/3}MnO_3$ is a compound where spontaneous phase separation plays a crucial role. This compound is AFM/FM phase segregated at low temperatures, while existing as a paramagnetic single phase at room temperature (Fig. 1) [29, 30, 31]. Below room temperature $Nd_{2/3}Ca_{1/3}MnO_3$ exhibits a sequence of phase transformations.

Thus, in this compound a first-order martensitic type, charge-ordering phase transformation takes place at $T_{CO} \approx 212$ K [32]. This results in coexistence of self-organized charge-ordered (CO) and charge-disordered (CD) phases in a wide temperature range, as evidenced by an extended temperature hysteresis of the magnetic susceptibility in the CO region. Below $T_{CO}$ the compound exhibits two antiferromagnetic transformations at $T_{N1} \sim 130$ K and $T_{N2} \sim 80$ K, and a ferromagnetic one at $T_C \sim 70$ K [29, 30]. Therefore, at low temperatures below about 70 K at least three different magnetic phases, including two AFM charge-ordered phases and a FM charge-disordered one, coexist. This low temperature magnetic phase-segregated state develops from the CO–CD phase segregated state, so that the CO phase becomes the AFM phase, and the CD phase transforms into the FM phase. At 4 K the AFM phase fraction totals to about 82% of the compound volume, and the FM phase fraction to about 18% [31]. The low temperature AFM/FM phase segregated state of $Nd_{2/3}Ca_{1/3}MnO_3$ single crystals has been observed directly using a scanning superconducting quantum interference device-based (SQUID) microscope [33].

Magnetic behavior of $Nd_{2/3}Ca_{1/3}MnO_3$ compound is similar to that of a cluster-glass magnetic state below the freezing temperature $T_g \sim 60$ K, which is close to the Curie temperature $T_C \sim 70$ K [34]. The structure and magnetic properties of the phase segregated state of this compound suggest the existence of the exchange bias effect due to the interface AFM/FM exchange coupling. In this work effects of temperature and magnetic fields on EB in a spontaneously phase segregated $Nd_{2/3}Ca_{1/3}MnO_3$ are studied experimentally in detail.

## 2. Experiment

Polycrystalline bulk $Nd_{2/3}Ca_{1/3}MnO_3$ samples were prepared by a standard solid state reaction technique from stoichiometric amounts of $Nd_2O_3$, $CaCO_3$ and $Mn_2O_3$ powders. After prefiring at 900°C, the mixture was pressed in the form of tablets and sintered at 1500°C for 10 h. Then the tablets were cooled slowly with a furnace at a rate of 80°C $h^{-1}$. X-ray - analysis of the crystal structure of the obtained compound materials indicated that the material was in a single-phase state at room temperature. At the same time, a neutron diffraction study has shown that at room temperature the crystal structure of the compound corresponds to the standard orthorhombic *Pnma* space symmetry with lattice parameters $a = 5.45$ Å, $b = 7.63$ Å, and $c = 5.39$ Å [29]. This finding agrees well with previously reported data [35].

Magnetic measurements were performed using Quantum Design Magnetic Properties Measurement System (MPMS) and a noncommercial SQUID magnetometer. Magnetic hysteresis loops were observed and characterized within a temperature range from 10 K to 80 K for $Nd_{2/3}Ca_{1/3}MnO_3$ samples (1) cooled in the zero magnetic field and (2) cooled in applied magnetic fields $H_{cool}$ up to 50 kOe. After each measurement of hysteresis loops of the total magnetization $M(H)$ a studied sample was demagnetized by warming up to 320 K and keeping it at 320 K for 1800 sec.

## 3. Results and discussion

### 3.1. Exchange bias effect

Considering the structure and magnetic properties of the spontaneously AFM – FM phase segregated state of $Nd_{2/3}Ca_{1/3}MnO_3$ compound at low temperatures (Fig. 1) [29] that suggests exchange coupling between the FM clusters and the AFM matrix, the exchange bias effect was expected in this compound. Thus, *M(H)* hysteresis loops were measured for several $Nd_{2/3}Ca_{1/3}MnO_3$ samples at different temperatures, and at both zero field cooled (ZFC) and magnetic field cooled (FC) conditions. For the ZFC experiments, a sample was cooled in the zero magnetic field from room temperature to temperatures in the range from 10 K to 75 K. For the FC experiments, a sample was cooled in different magnetic fields (from $H_{cool} = 0.8$ to $H_{cool} = 50$ kOe), while temperature was changed from room temperature to temperatures in the range from 10 K to 75 K.

A magnetic field induced shift of a hysteresis loop of *M(H)* that defines the exchange bias effect is determined as

$$H_{EB} = (H_1 + H_2)/2, \qquad\qquad (1)$$

and the coercive field as

$$H_C = (H_2 - H_1)/2, \qquad\qquad (2)$$

where $H_1$ and $H_2$ are magnetic fields corresponding to zero magnetization (see the insert in Fig. 2 ).

The obtained experimental data revealed an important property of the measured magnetization loops. Thus, all ZFC loops were centered with regard to the intersection of the zero field and zero magnetization axes (Fig. 2), while FC loops were shifted both toward the negative direction of the magnetic field axis and the positive direction of the magnetization axis. The observed exchange bias effect signifies the presence of unidirectional exchange interaction anisotropy driving the magnetic moments of FM clusters back to their initial orientation when the magnetic field is removed. In particular, when a sample is cooled down to 10 K in the magnetic field $H_{cool} = 20$ kOe (Fig. 2), the total magnetic moment reaches its negative values at $H_1 \sim$ - 0.9 kOe, while the magnetic field decreases from its maximum value (20 kOe; decreasing branch of the loop). In the case of the increasing branch of the loop, the total magnetization *M(H)* changes sign at $H_2 \sim 0.5$ kOe. Using these data in Eqs. (1) and (2) one can calculate values of the exchange bias field $H_{EB} \sim$ - 0.3 kOe and the coercive field $H_c \sim 0.7$ kOe, respectively.

Evolution of the exchange bias effect with cooling magnetic field can be demonstrated measuring magnetization loops at 10 K (Fig. 3) for a sample cooled in different $H_{cool}$. The obtained results for $H_{EB}(H_{cool})$ are discussed in Sec. 3.3.

It has to be noted, that in low magnetic fields ($H_{cool} \leq 1$ kOe) FC hysteresis loops are shifted along the magnetization axis toward its positive direction, because the hysteresis loops measured between $\pm H_{cool}$ are minor hysteresis loops. An increase in the cooling magnetic field and the corresponding increase in the measuring magnetic field lead to a decrease in the vertical shift of the loops, as they approach the full hysteresis loop (Fig. 3b, Fig. 2).

An important property of materials with exchange bias is a training effect observed in continuously field-cycled [20, 26] materials. The $M(H)$ loops measured for magnetic fields in the range from 1 to $-1$ kOe after field-cooling a sample in the field $H_{cool} = 1$ kOe are shown in Fig. 3 a for the loops with the index number $\lambda = 1$, 2, and 3, where the value of $\lambda$ denotes the number of cooling cycles. The training effect is manifested through a decrease in the exchange bias effect and the coercive field value with an increase in the loop index number. These decreases are caused by relaxation of the sample material to offset exchange interaction anisotropy that leads to asymmetry of the remanence. Such asymmetry is particularly significant between the first and second cycles, as evidenced by the $M(H)$ curves for $\lambda = 1$ and 2 in Fig. 3a. A physical mechanism behind the training effect is relaxation of surface spin configurations of FM clusters' spins and those of the AFM matrix toward a local equilibrium configuration [20].

Yet another interesting feature of the FC hysteresis loop with the index number $\lambda = 1$ in Fig. 3 a is its openness. Similar unclosed loops of the total magnetization were observed for other systems exhibiting exchange bias. They are a result of relaxation of a nonequilibrium magnetic state of the corresponding FM cluster subsystem [36, 20] to some local equilibrium state. With an increase in strength of external magnetic fields over 20 kOe such hysteresis loops become closed, and the training effect is less pronounced. This happens due to suppression of the glassy magnetic state [26], as non-equilibrium FM cluster subsystems relax to the global equilibrium. In figures presented in this paper all data (except some of those in Fig. 3) are shown for the loops with the index number $\lambda = 1$.

A temperature dependence of the EB phenomenon is demonstrated in Fig. 4. Here, $M(H)$ loops specific to the case of the cooling magnetic field $H_{cool} = 0.8$ kOe and measured in external magnetic fields from 0.8 kOe to $-0.8$ kOe are shown for temperature values a) 20 K, b) 35 K, and c) 75 K. These data indicate that a decrease in the exchange bias field $H_{EB}$ with an increase in temperature is associated with a decrease in the volume fraction of the FM phase in the studied sample accompanied by a reduction in the total FM/AFM interface area. These findings are supported by previous neutron diffraction data [29] of the authors. Correspondingly, a decrease in the coercive force with increasing temperature follows from a decrease of the total volume of FM nanodomains (Fig. 4 a, b). Above $T_C$ temperature FM nanodomains disintegrate, and the EB effect vanishes (Fig. 4 c), as demonstrated by coinciding ZFC and FC loops.

### 3.2. Characterization of the low temperature ferromagnetic phase

In the case of the low temperature phase segregated state of the studied $Nd_{2/3}Ca_{1/3}MnO_3$ compound it is important to separate FM and AFM subsystem contributions to the total magnetization of the system.

At temperatures above $T_C \sim 70$ K $M(H)$ curves are non-hysteretic and show a nearly linear dependence of the total magnetization on the magnetic field (Fig. 4 c), thus indicating that FM phase is absent. In contrast, below $T_C$ $M(H)$ curves are hysteretic and bend at higher magnetic fields, indicating a significant FM phase contribution to the total magnetization (Fig. 4 b). At the same time, these curves do not appear to approach the saturation value of the total magnetization as the field approaches 50 kOe. Instead, the total magnetization continues to grow as a nearly linear function of the magnetic field in high magnetic fields up to 90 kOe [37]. This behavior of the total magnetization curves indicates a significant contribution of the AFM phase [37] to the total magnetization.

To evaluate the AFM contribution to the total magnetization, experimental data for $M(H)$ taken at $T$=10º K in relatively high magnetic fields $3.2 < H < 5$ kOe were approximated by a sum of FM and AFM contributions: $M(H) = M_s \cdot (1 - c_1/H) + \chi_a \cdot H$ ,

where $M_s \cdot (1 - c_1/H)$ is the FM contribution obtained in a high-field approximation, and $\chi_a \cdot H$ is the AFM one. All three quantities $M_s$, $\chi_a$, and $c_1$ are fitting parameters dependent on temperature, with $M_s$ signifying the saturation magnetization and $\chi_a$ the AFM susceptibility. To confirm validity of the above linear fitting of the AFM contribution, the exact same fitting was also applied to experimental $M(H)$ data taken at $T$=10º K in strong magnetic fields up to 90 kOe reported earlier in Ref. 37 for the same compound. For the data of Ref. 37 the obtained high-field value of $\chi_a$ is about 0.08 emu/kOe. This value agrees reasonably well with the value 0.11 emu/kOe obtained by fitting experimental data of this study.

A similar procedure for estimating the AFM contribution was performed for different temperatures below 70 K. Then the contribution of the FM phase, $M_{FM}(H)$, to the total magnetization was obtained by subtraction of the fitted AFM one from the experimental values of the total magnetization (Fig. 5).

The exchange bias effect is clearly manifested as a shift of the FM phase magnetization loops (see the insert in Fig. 5). It can also be seen that the coercive field specific to $H_{cool} = 50$ kOe is much higher than that specific to $H_{cool} = 10$ kOe  or 20 kOe. This dependence of the coercive field on the cooling field is associated with AFM to FM transformation occurring in magnetic fields above about 30 kOe [35] that results in continuing growth of the volume fraction of the FM phase at the expense of the surrounding AFM matrix.

### 3.3. Temperature and cooling field dependence of EB effect

Dependence of the EB field on the cooling magnetic field, $H_{EB}(H_{cool})$, at 10 K, and the EB field dependence on temperature, $H_{EB}(T)$, in the cooling field $H_{cool} = 0.8$ kOe, obtained from $M_{FM}(H)$ loops are presented in Figs. 6 and 7, respectively.

At a constant temperature, $H_{EB}(H_{cool})$ is a strongly non-monotonous function of the cooling field, while for the zero cooling field $H_{EB}$ is zero. For the cooling magnetic field of about 1 kOe the function $H_{EB}(H_{cool})$ has a sharp minimum (Fig. 6). Most likely, this minimum value is due to the fact that for low values of $H_{cool}$ only that part of the FM clusters population, for which the condition $H_c < H_{cool}$ is fulfilled, is orientated along the external magnetic field. Consequently, only this portion of the FM cluster subsystem contributes to the EB effect.

With the further increase in the cooling field $H_{EB}$ decreases. In particular, $H_{EB}$ drops to about 75% of its value at $H_{cool} = 2$ kOe when $H_{cool}$ reaches 50 kOe. These data are in a good agreement with experimental results for a spontaneously phase separated cobaltite $Pr_{0.7}Sr_{0.3}CoO_3$ reported in Ref. 26, where it was shown that the decrease of $H_{EB}$ could be associated with growth of the FM clusters. These data seem to be also consistent with the Meiklejohn model evaluation of $H_{EB}$ in a two layer AFM - FM system [8], $H_{EB} \approx J_{ex} / \left( M_{FM} \cdot t_{FM} \right)$, where $J_{ex}$ is the FM/AFM exchange interaction constant per unit area, and $M_{FM}$ and $t_{FM}$ are the magnetization and thickness of the FM layer, respectively. However, one has to take into account that EB mechanisms acting in the phase segregated compounds and those in thin films composed of two coupled AFM and FM layers may differ significantly. This is primarily due to the fact, that the spontaneously phase segregated compounds possess a lamellar structure [38], where each FM plate is bordered by two AFM layers, thus providing for an AFM – FM –AFM – type exchange coupling.

Similarly, the decrease in $H_{EB}$ with increasing $H_{cool}$ in phase segregated compounds is likely to be associated with a suppression of their glassy magnetic state [39].

The $H_{EB}(H_{cool})$ curves obtained in the present study similar in nature to those specific to phase separated cobaltites $La_{0.88}Sr_{0.12}CoO_3$ [24] and $Pr_{0.7}Sr_{0.3}CoO_3$ [26], and manganite films [40]. Both $H_{EB}(T)$ and $H_c(T)$ fields measured at the fixed cooling magnetic field $H_{cool} = 0.8$ kOe (Fig. 7) reach their respective maximum values at low temperatures. With increasing temperature these fields sharply decrease and vanish above $T_C \sim 70$ K, where FM phase nanodomains transform to become paramagnetic charge-disordered clusters coexisting with the charge-ordered AFM matrix. These observations are in a good agreement with earlier neutron diffraction data of the authors demonstrating that the FM component of the total magnetic moment gradually grows when temperature decreases below 70 K [29].

The experimental studies presented here prove that the exchange bias effect strongly depends on temperature and magnetic field. This dependence is directly related to phase separations in the studied $Nd_{2/3}Ca_{1/3}MnO_3$ perovskite manganite. The obtained results indicate that EB-related properties of spontaneously phase segregated compounds may be tuned by manipulations with temperature and cooling magnetic fields.

In a segregated phase state, the studied $Nd_{2/3}Ca_{1/3}MnO_3$ compound consists of FM nanoclusters immersed in AFM matrix. Therefore, to further understand the nature of exchange bias in such compounds systematic experimental and theoretical studies of exchange bias in nanoscale systems are necessary. The first results already obtained in the course of fundamental theory-based computations and modeling of small quantum dots [41, 42] provide an insight into physical and chemical mechanisms of exchange interaction effects leading to the exchange bias development and loss. Similar studies of exchange bias in $Nd_{2/3}Ca_{1/3}MnO_3$ nanosystems are paramount to explain and manipulate with exchange bias in this and other phase segregated compounds.

## 4. Conclusions

Exchange coupling observed in $Nd_{2/3}Ca_{1/3}MnO_3$ compound, which is intrinsic to this system, manifests the existence of a low temperature spontaneously AFM - FM phase separated state of this material. A dependence of the exchange bias field in this compound on temperature and the cooling magnetic field was discovered and characterized by studying samples of this compound cooled through both the Néel and the Curie temperatures in magnetic fields from below 1 kOe to over 50 kOe. The corresponding training effect that supports experimental evidences of the existence of exchange bias in the studied material has also been observed and evaluated. The exchange bias effect exhibits a strong dependence on cooling magnetic field and temperature caused by evolution of the FM phase content and the FM clusters size in the phase segregated state of the compound.

The FM phase contribution to the total magnetization of the system has been evaluated. The exchange bias effect vanishes above the Curie temperature with disappearance of FM phase clusters.

## 5. Acknowledgements

Authors are grateful to Dr. D. Khalyavin for fruitful collaboration. This work was supported by Slovak Research and Development Agency through the contract No. APVV-0132-11, and by the State Fund of NAS, Ukraine, through the contact No.4/11-H.

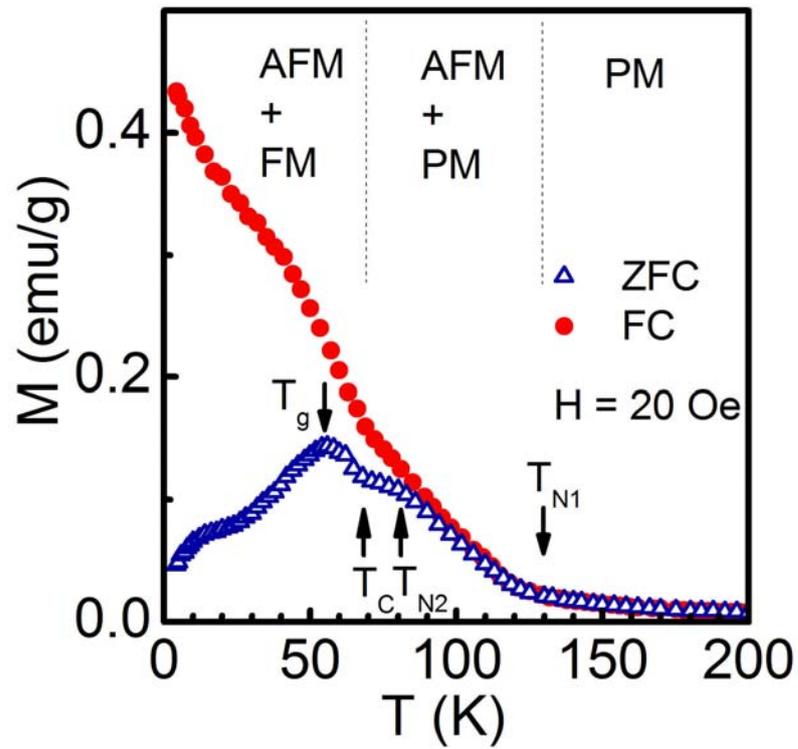

Fig. 1. (Color online) ZFC (full symbols) and FC (open symbols) static magnetizations of $Nd_{2/3}Ca_{1/3}MnO_3$ compound as a function of temperature measured in the magnetic field $H = 20$ Oe. Dashed lines indicate the temperature values corresponding to transitions between different magnetic states; PM denotes a paramagnetic phase, AFM + PM coexisting antiferromagnetic and paramagnetic phases, and AFM + FM coexisting antiferromagnetic and ferromagnetic phases, respectively.

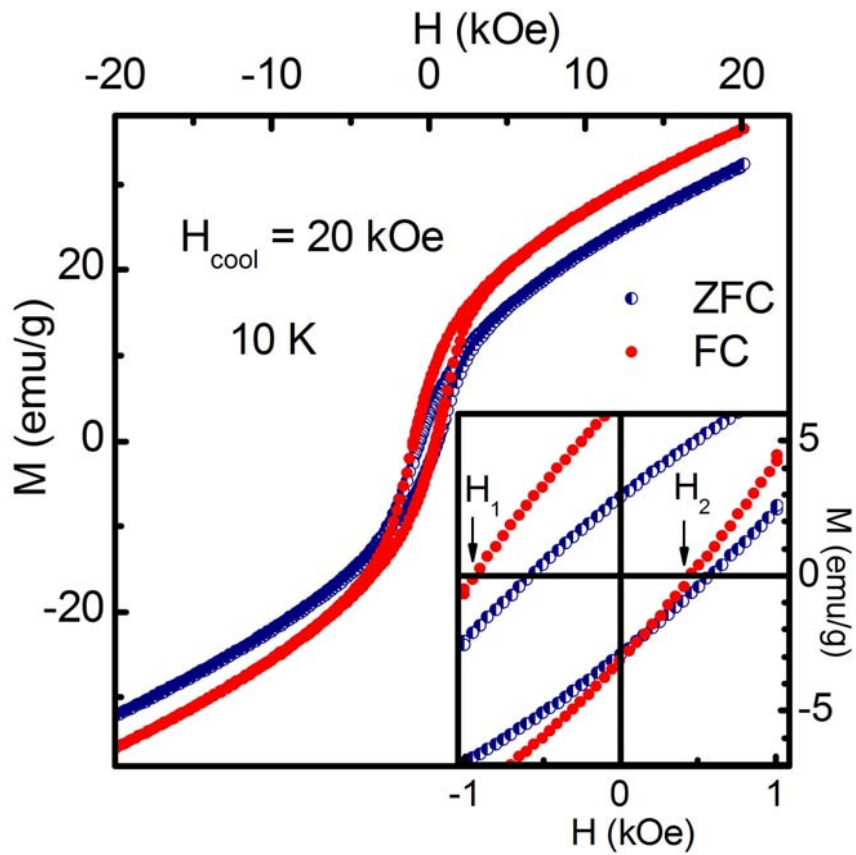

Fig. 2. (Color online). Hysteresis loops of the total magnetization specific to $Nd_{2/3}Ca_{1/3}MnO_3$ compound at 10 K measured after zero-field-cooling and field-cooling ($H_{cool}$ = 20 kOe). Insert: an enlarged view of the central region of the loops measured at 10 K.

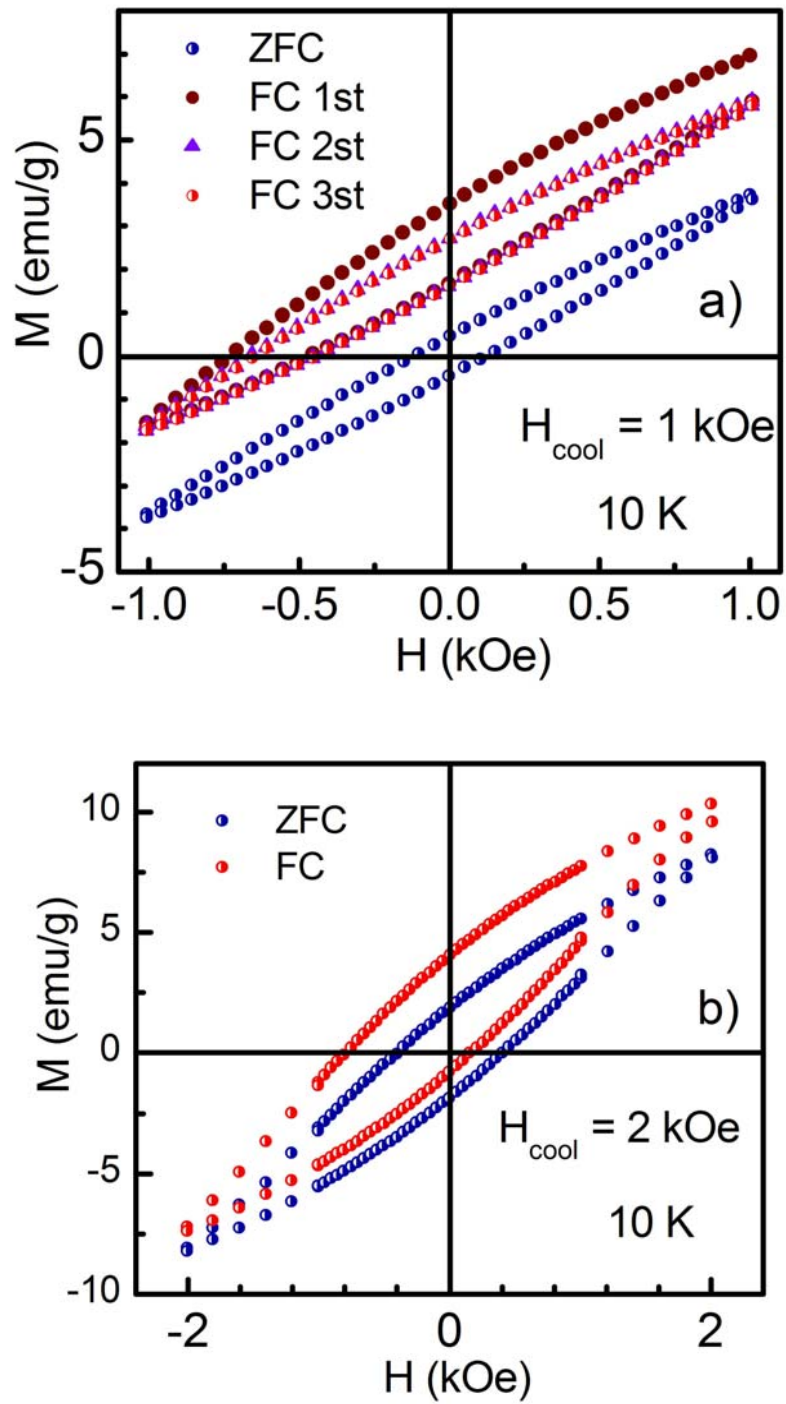

Fig. 3. (Color online). Hysteresis loops of the total magnetization specific to $Nd_{2/3}Ca_{1/3}MnO_3$ compound at 10 K measured after zero-field-cooling and field-cooling in magnetic fields a) $H_{cool}$ =1 kOe and b) $H_{cool}$ = 2 kOe.

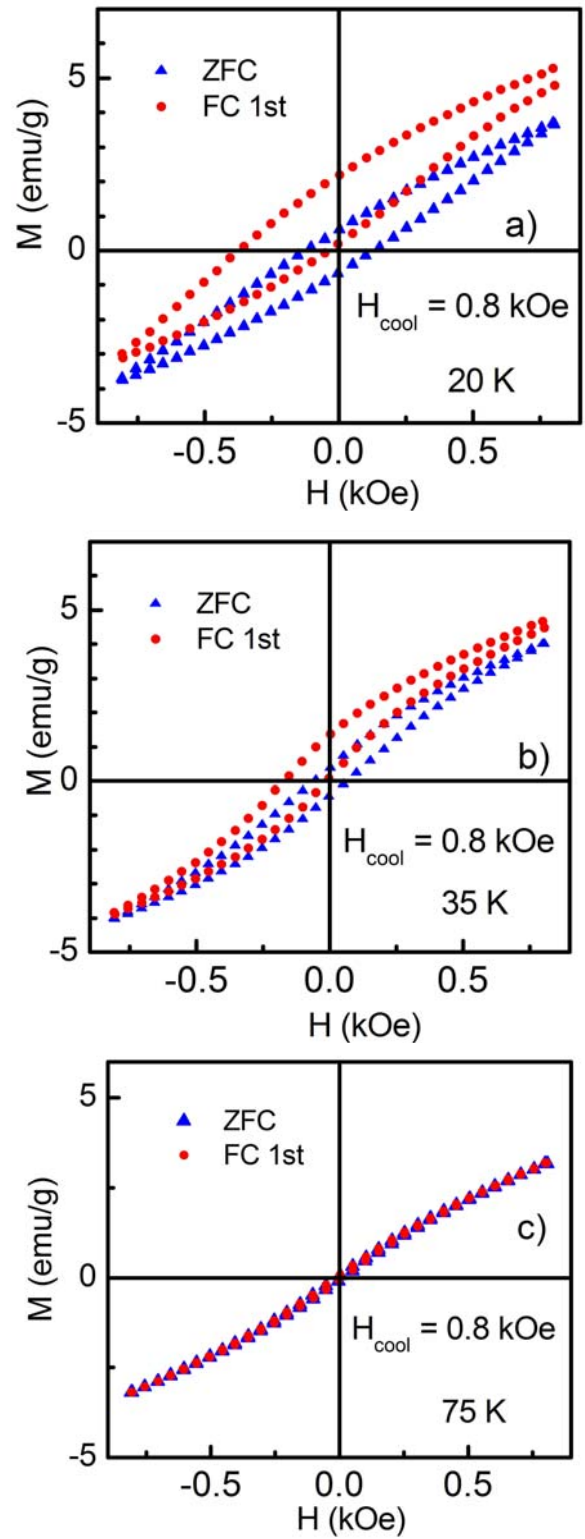

Fig. 4. (Color online). Hysteresis loops of the total magnetization specific to $Nd_{2/3}Ca_{1/3}MnO_3$ compound measured after zero-field-cooling and field-cooling in the magnetic field $H_{cool}$ =0.8 kOe at different temperatures a) 20 K, b) 30 K, and c) 75 K.

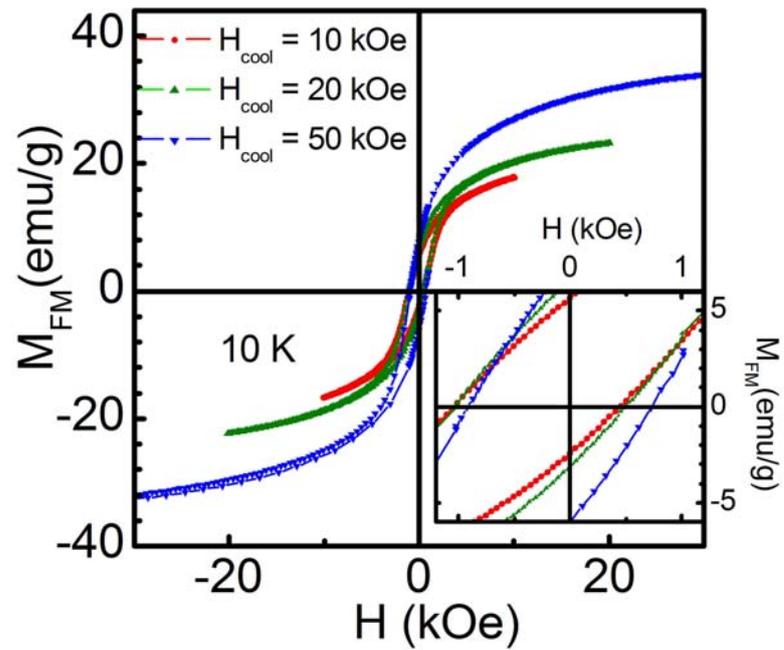

Fig. 5. (Color online). Loops of the FM contribution to the total magnetization, $M_{FM}(H)$, measured after zero-field-cooling and field-cooling ($H_{cool}$ = 10 kOe, 20 kOe, and 50 kOe), and calculated by subtracting a contribution due to the AFM matrix from the total magnetization values ($T$ = 10 K, $H_{cool}$ = 10 kOe).

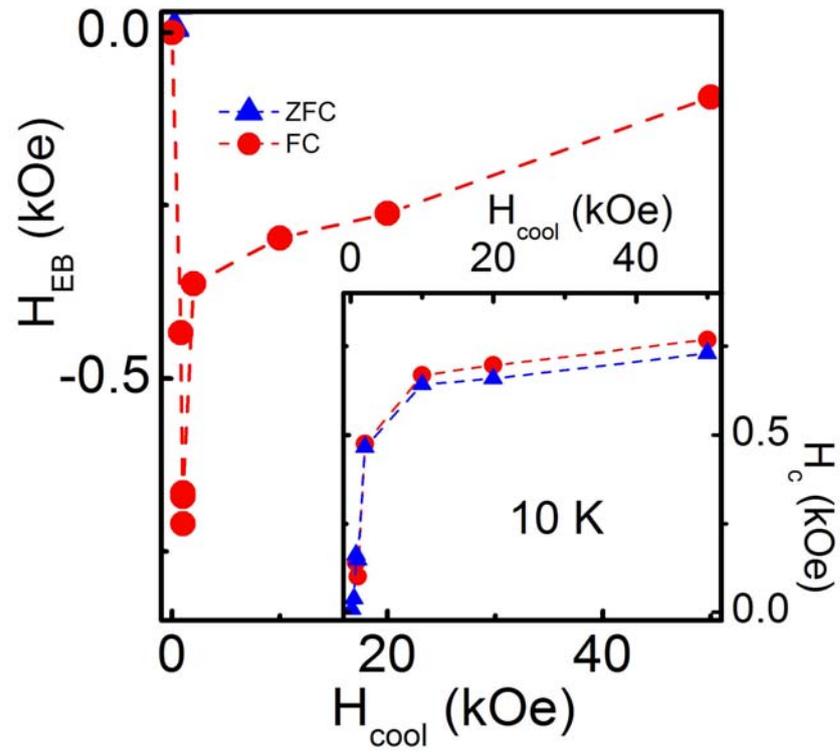

Fig. 6. Cooling field dependence of the exchange bias field $H_{EB}(H_{cool})$ at 10 K for $Nd_{2/3}Ca_{1/3}MnO_3$ compound. Lines are guides for an eye only. Insert: a cooling field dependence of the coercive field $H_c(H_{cool})$.

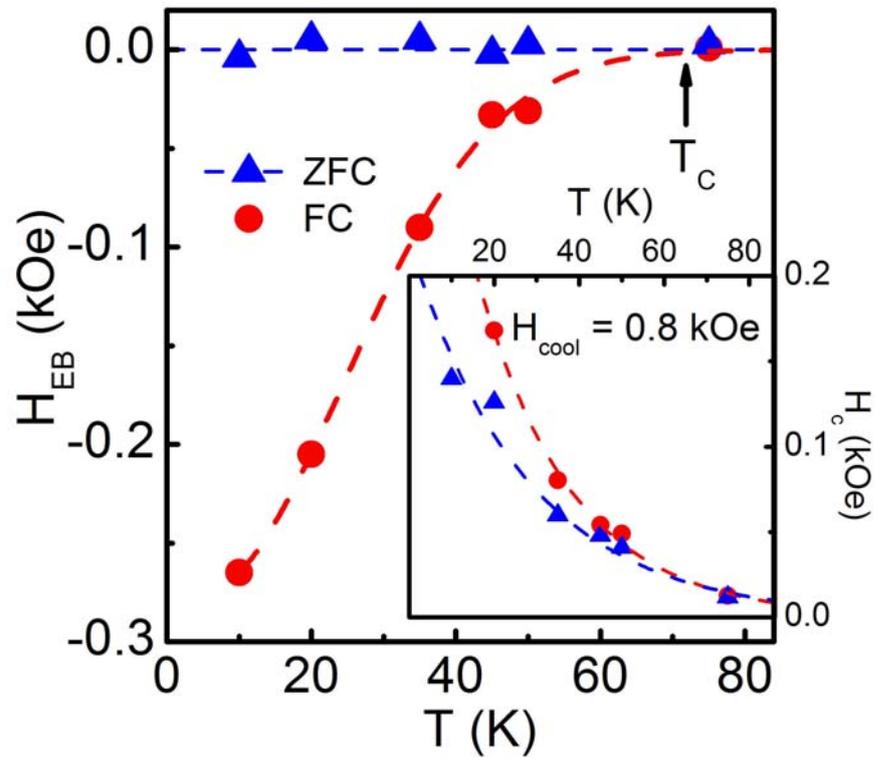

Fig. 7. (Color online) Temperature dependence of the exchange bias field $H_{EB}(T)$ for Nd$_{2/3}$Ca$_{1/3}$MnO$_3$ compound measured after field cooling ($H_{cool}$ = 0.8 kOe) and zero-field cooling. Lines are guides for an eye only. Insert: a temperature dependence of the coercive field $H_c(T)$ as a function of temperature.